\documentclass[english]{article}
\usepackage[T1]{fontenc}
\usepackage[latin9]{inputenc}
\usepackage{geometry}
\usepackage{color}
\usepackage{graphicx}
\usepackage{verbatim}
\usepackage{amssymb}
\usepackage{epstopdf}
\usepackage{amsmath}
\usepackage{mathtools}
\usepackage{caption}
\usepackage[normalem]{ulem}
\usepackage[makeroom]{cancel}
\usepackage{cite}
\usepackage{amsmath}
\usepackage[toc]{appendix}
\usepackage[colorlinks,citecolor=magenta,linkbordercolor={0 0 1}]{hyperref}

\newcommand{\comments}[1]{}   

\makeatletter

\usepackage{indentfirst}
\usepackage{enumerate}

\makeatother

\begin{document}

\title{Typicality from concentration of measure in models with binary state variables}

\author{Nicol\'as Nessi$^1$\\
$^1$ IFLP CONICET,\\
Diagonal 113 y 64, La Plata, Buenos Aires, Argentina.
}

\maketitle

\abstract{
We study the emergence of typicality in classical systems with a large number of binary state variables. We show analytically that for sufficiently large subsets of the complete state space, state functions which can be associated with macroscopic observables, such as density or energy, are sharply concentrated around a typical value, i.e., the vast majority of microscopic states show the same macroscopic behavior. From this static typicality result we obtain the dynamical counterpart, which states that if two initial conditions are drawn from a sufficiently large subset of the state space, the dynamical trajectories of the macroscopic observables would be approximately the same, provided that the dynamics does not compress abruptly the size of sets of states under evolution. The ensuing dynamical typicality phenomenon is very similar to what was recently found in the context of the dynamics of isolated quantum systems. We illustrate and apply our analytical results by analyzing one-dimensional cellular automata.
}



\section{Introduction}
\label{sec:introduction}

We speak of \textit{typicality} whenever we find that in a certain set of states of a system, grouped according to some common feature (drawn with the same probability distribution, sharing the same value of some relevant observable, etc.), the vast majority of the members of the set share (at least approximately) some other feature. Typicality and \textit{inference} are often related. In fact, if the vast majority of the members of one set share similar (typical) features, it is possible to infer with high confidence the features of a yet unknown member of the set by knowing the features of another. It is in this sense, that the presence of typicality provides predictive power to the conclusions extracted by \textit{inference}.

To illustrate the concept of typicality and to put into perspective its importance, we will take classical statistical mechanics as an example. Among the several approaches to the foundations of classical statistical mechanics, the subjective approach proposed by Jaynes is distinctive~\cite{jaynes57_info_stat_mech_I}, since it is not based on dynamical considerations, at variance with other approaches like ergodic theory or Khinchin's thermodynamic limit approach~\cite{uffink06_compendium_found_stat_mech}. Instead, Jaynes showed, building on the equivalence between Shannon and thermodynamic entropies, that Gibbs' distribution can be considered as the probability measure that \textit{only} assumes the (average) value of the energy, since it maximizes ignorance (entropy) subject to such constraint. In other words, the maximum entropy distribution does not emerge from the dynamical properties of the system, but it is, instead, the best inference that we can make starting from the total energy of the system. It is typicality which underlies the predictive power of the maximum entropy distribution, which, in the end, is just the broadest possible distribution compatible with the information available. In Jaynes' words~\cite{jaynes57_info_stat_mech_I} (emphasis in original):
\begin{quote}
Phenomena in which the predictions of statistical mechanics is well verified experimentally are always those in which our probability distributions, for the macroscopic quantities actually measured, have enormously sharp peaks. (...) Evidently, such sharp distributions for macroscopic quantities can emerge only if it is true that for \textit{each} of the overwhelming majority of those states to which appreciable weight is assigned, we would have the \textit{same} macroscopic behavior. We regard this, not merely as an interesting side remark, but as the essential fact without which statistical mechanics could have no experimental validity, and indeed without which matter would have no definite macroscopic properties, and experimental physics would be impossible.
\end{quote}

The use of typicality arguments to establish the foundations of statistical mechanics was also pursued recently in the context of quantum systems. Concretely, it was discovered that for the overwhelming majority of the wave functions of a sufficiently large system which lie inside a microcanonical window of energies, the reduced density matrix of any subsystem is well approximated by a canonical density matrix, a feature designated as ``canonical typicality''~\cite{goldstein06_canonical_typicality,popescu06_canonical_typicality}. In Ref.~\citeonline{reimann07_typicality_microcanonical} it is demonstrated that states drawn according to a certain probability distribution in Hilbert space feature very similar quantum expectation values of generic observables.

It is interesting that the concept of typicality is of relevance also for the statistical description of \textit{dynamical processes} and not only states. Again, using a quotation by Jaynes~\cite{jaynes85_macroscopic_prediction}:
\begin{quote}
... to predict the course of a time-dependent macroscopic process, choose that behavior that can happen in the greatest number of ways, while agreeing with whatever information you have - macroscopic or microscopic, equilibrium or nonequilibrium.
\end{quote}
In other words, when studying a macroscopic process, it is more likely to observe a typical realization of the dynamics, the behavior that can happen in the greatest number of ways. This may sound as a tautology, but the fundamental content of the statement is that such typical behavior does exist. The absence of a typical behavior would imply the impossibility to reproduce macroscopic dynamical phenomena, because any small difference in the preparation of the initial condition or some other factor would lead to a different macroscopic behavior. As Jaynes puts it in Ref.~\citeonline{jaynes85_macroscopic_prediction}:
\begin{quote}
If any macrophenomenon is found to be reproducible, then it follows that all microscopic details that were not reproduced, must be irrelevant for understanding and predicting it.
\end{quote}

In fact, typicality was shown to be an ubiquitous property of the \textit{dynamics} of isolated quantum systems.  In a pioneering investigation, Bartsch and Gemmer demonstrated that the overwhelming majority of pure states in a set featuring a common expectation value of some observable $A$ at some time $t$, yield very similar expectation values at any later time~\cite{bartsch09_dyn_typ_first}. In Ref.~\citeonline{reimann20_dyn_typ} it was shown that the vast majority of pure states with energies inside a microcanonical interval which at time $t=0$ have very similar values of some set of observables $A_1, \ldots,A_K$ will exhibit very similar expectation values for \emph{any} other observable $O$ at any later time. The reproducibility of macroscopic experiments in quantum systems rests upon this remarkable feature~\cite{reimann20_dyn_typ}.

Moreover, equilibrium and dynamical typicality in quantum systems are intimately related. In fact, equilibrium and dynamical typicality in quantum systems can be considered as a manifestation of the \textit{concentration of measure phenomenon}~\cite{talagrand96_new_look_independence,ledoux05_the_com_phenomenon} on the high dimensional sphere~\cite{popescu06_canonical_typicality,reimann20_dyn_typ}.

In this paper, we study typicality in what is perhaps the simplest \textit{classical} system, a collection of entities with only two states. We are able to extract analytical results in this context. Our results have implications for the static as well the dynamic behavior of large collections of binary variables.
The general message which emerges is that equilibrium and dynamical typicality in this class of systems are also related, and, moreover, they are also a consequence of a concentration phenomenon. This time, the relevant space is the Boolean cube, the set of all possible states of a collection of binary state variables.

Concentration of measure of Lipschitz functions on the \textit{whole} Boolean cube is a well known phenomenon in the mathematical literature~\cite{barvinok05_com}\cite{root10_phd_thesis_com_boolean_cube}. In order to apply these results in physical scenarios, we generalize them to the case of subsets of the whole state space. This is because in physical settings, one is not interested in the whole space of possible states of a system, but only in a subset of states, usually selected according to some restriction. The typical example would be a microcanonical window of energies, composed by states with approximately the same energy.

In particular, we are able to show that for any \textit{sufficiently large} set of states, the value of all state functions corresponding to macroscopic observables, i.e., observables receiving contributions from all variables, concentrate around the median inside the set, provided that the number of variables is large. In other words, these observables exhibit typical values inside the given set.
From the static, or equilibrium typicality result is possible to extract a result pertaining the time evolution of sets of states under discrete dynamics. Basically, if we start with a sufficiently large set of states in which suitable observables are concentrated, and if the dynamics is such that there are no massive merging of states, i.e., the dynamics does not shrink abruptly the size of the set, then, at any finite time, concentration of observables will persist. This constitutes an instance of dynamical typicality in a classical system.

We illustrate the analytical results through an application to one-dimensional cellular automata (CA), which are among the simplest dynamical systems conceivable. CA are dynamical systems in which space, time and states are discretized, and the dynamical evolution is implemented through uniform local rules that update the state of the system at each time step. Despite their simplicity, CA have are appealing because they can be considered as models of a wide variety of complex systems, like fluids, neural networks, ecologic and economic systems, chemical reaction-diffusion systems, crystal growth, pattern formation in living systems, traffic flow, urban segregation, etc.~\cite{ilachinski01_ca_book,wolfram02_nks_book}. Moreover, they can be regarded, in abstract, as models of complexity itself, and constitute a playground to study emergent properties in complex systems in its most essential form~\cite{ilachinski01_ca_book,berto_ca_stanford_encyclopedia}. In Section~\ref{sec:application}
 we focus on the most simple CA, one-dimensional chains of sites with two states evolving with a nearest neighbor rule, called elementary CA.

The rest of the paper is organized as follows. In section~\ref{sec:intuition} we present a physically intuitive formulation of our results. In section~\ref{sec:rigour} we present a formulation of the results in a mathematically rigourous way, together with the corresponding proofs. In section~\ref{sec:application} we analyze the case of one dimensional CA using our analytical results, contrasting them with the numerical findings. Finally, in section~\ref{sec:conclusions}, we formulate our conclusions and outlook.

\section{Main results}
\label{sec:intuition}
In this section we will state our main results in a physically intuitive way. In the next section we will provide a rigourous formulation in the form of theorems. We are interested in the space of all possible configurations of a systems with $n$ binary variables, whose attributes we take to be $\{0,1\}$ for convenience. Such space, that we will denote as $I_n$, has dimension $2^n$. Our results will apply to systems living in lattices with arbitrary geometry and in any number of dimensions, or even in more general graphs, as long as the number of variables $n$ is large. This includes many fundamental models with binary variables, such as the classical Ising model and binary CA.

We are interested in what we will call \textit{macroscopic observables} on $I_n$. As a provisional and intuitive definition, we say that a function $f(x)$, where $x=(\xi_1,\xi_2,\cdots,\xi_n)$ with $\xi_i=0$ or $1$ is a state in $I_n$, is a macroscopic observable if it can be written as a sum of contributions coming from many of the $\xi$s that define $x$ without putting too much weight on any $\xi_i$ in particular. A typical example could be the density of sites in state $1$, $\rho(x)=\frac{1}{n}\sum^{n}_{i=1}\,\xi_i$. We will give a mathematical definition of macroscopic observable in the next section.

Our main result can be stated as follows:

\textbf{Typicality in $I_n$}: \textit{For any sufficiently large subset $A$ of $I_n$ with large $n$, all macroscopic observables have approximately the same value for the vast majority of states in $A$.}

Notice that it is not relevant how we choose the states that conform $A$ as long as the number of states is large enough. However, in general, we will be thinking about a set of states that satisfy a given constraint or set of constraints, which we will denote as $R$. For example, let us consider a microcanonical window of energies in the Ising model. Since the variables of the Ising model take values $\{1,-1\}$, in order to make connection with our setup, we should define the Ising variables as $S_i=2\xi_i-1$. We choose $A$ to be the set of all states whose Ising energy $E(x)=-J\sum^{n-1}_{i=1}\,S_i S_{i+1}$ is inside an interval $E_0-\Delta E\leq E(x)\leq E_0+\Delta E$ centered around $E_0$ with $\Delta E \ll E_0$. This defines a microcanonical ensemble with energy $E_0$. Our result indicates that all macroscopic observables will have approximately the same value for almost all states inside the microcanonical set, as long as this set is large enough. For example, our results will not apply in the case $E_0=-J(n-1)$, the ground state, since such microcanonical set has only two states, all spins up or all spins down, given a sufficiently small $\Delta E$.

Another important result follows from the previous statement:

\textbf{Dynamical typicality in $I_n$}: \textit{Consider that we form a set $A_R$ with states satisfying a given constraint $R$. We will use the states in $A_R$ as initial conditions, and make them evolve a given number of steps according to some discrete dynamical rule, which can be deterministic or stochastic. The statement is that for large $n$, and if the set $A_R$ is large enough, the value of any macroscopic observable will be very similar for most of those initial states at any time (including the initial time), provided that the dynamics does not compress abruptly the set of states}.

What do we mean with the dynamics compressing the set of states? This is related with the measure of a certain set, something that we will introduce in the next section. For the moment, let us say that the size of a \textit{group} of states is proportional to the number of \textit{different} states in it. We will define a \textit{set} as any group with no repeated states. We will consider that the initial group of states $A$ is indeed a set. Although the initial set has no repeated states, after a given number of steps, the resulting group of states may have repeated states because the dynamical rule can map two different states at time $t$ into the same state at time $t+1$. Then ,the set of non-repeated states at a finite time will be smaller that the size of the initial set. That is what we refer as the dynamics shrinking the size of the set of states. If the dynamics is reversible, for example, a reversible cellular automata~\cite{takesue89_reversible_ECA}, the size of the set does not change, any state has only one predecessor and only one descendant. But there are dynamical rules, like some elementary CA or the stochastic dynamics of the Ising model that will eventually compress the set of states. The dynamical typicality statement will hold if this compressing does not reduces abruptly the size of the set of states.

\section{Concentration of measure in subsets of the Boolean cube}
\label{sec:rigour}

\subsection{Definitions and theorems}

In this section we will formalize the results that were presented in the previous section. In order to do that we need to introduce several mathematical definitions.

First, we will introduce a distance in $I_n$. We will define the Hamming distance as the number of bits whose values differ when comparing two different states $x=\left(\xi_1,\xi_2,\cdots,\xi_n\right)$ and $y=\left(\eta_1,\eta_2,\cdots,\eta_n\right)$,
\begin{equation}
\textrm{dist}\left(x,y\right)=\vert i\,:\xi_i\neq\eta_i \vert=\sum^n_{i=1}\,\vert \xi_i - \eta_i \vert,
\end{equation}
where $\vert A \vert$ denotes the number of elements in set $A$ and $\vert r \vert$ the absolute value of a real number $r$.

As we will be making statements about the size of subsets of $I_n$, the second important ingredient will be a measure on such a space, that we will take to be the counting (uniform) measure $\mu_n$, defined as,
\begin{equation}
\mu_n\{A\}=\frac{\vert A\vert}{2^n},
\end{equation}
i.e., the measure of a given set is taken to be the number of different states in the set normalized to the total number of states.

The third important ingredient will be the observables, that will be given by sufficiently well behaved functions $f(x)$ of the state of the system. In concrete, we will work with Lipschitz functions, whose rate of change is bounded according to,
\begin{equation}
\vert f(x)-f(y)\vert\leq\,\kappa\,\textrm{dist}\left(x,y\right),\,\forall\,x,y\in I_n,
\end{equation}
where $\kappa$, the Lipschitz constant, is a positive real number. We will denote as $\kappa^*$ the smallest value that $\kappa$ can take which fulfills the above inequality.

We will be interested in the distribution of values of $f(x)$ inside a given subset $A$ of $I_n$. To study this distribution we will use its median $m_f^A$, defined through,
\begin{eqnarray}
\mu_n\{x\in A : f(x)>m_f^A \}&\geq&\frac{\mu_n\{A\}}{2},\\
\mu_n\{x\in A : f(x)<m_f^A \}&\geq&\frac{\mu_n\{A\}}{2}.
\end{eqnarray}

With these definitions, the main theorem can be stated as follows:

\textbf{Theorem} \textit{Let $A\subset I_n$ be a non-empty set. Let $f:I_n\rightarrow \mathbb{R}$ be a Lipschitz function and $m^A_f$ its median inside $A$. Then, for all $\epsilon>0$},
\begin{equation}
\label{eq:main_result}
\mu_n\{x\in A\,:\,\vert f(x)-m_f^A \vert\geq\,\kappa\epsilon\sqrt{n}\}\leq\frac{4e^{-\epsilon^2}}{\mu_n\{A\}},
\end{equation}
\textit{or, conversely,}
\begin{equation}
\mu_n\{x\in A\,:\,\vert f(x)-m_f^A \vert\leq\,\kappa\epsilon\sqrt{n}\}\geq\mu_n\{A\}-\frac{4e^{-\epsilon^2}}{\mu_n\{A\}}.
\end{equation}

The theorem above is an example of what is known as a concentration of measure inequality in the mathematical literature, a bound on the measure of the set of states which deviate beyond some threshold from the reference value, usually the median or the mean. The space $I_n$, known as the Boolean cube, is one of the simplest spaces showing concentration phenomena~\cite{ledoux05_the_com_phenomenon,barvinok05_com,root10_phd_thesis_com_boolean_cube}. A similar result, but valid only for $A=I_n$ is a standard result on the Boolean cube~\cite{barvinok05_com}.


Now we will state two corollaries which establish sufficient conditions for an observable $f(x)$ to show concentration for large $n$. In this sense, they define the class of ``macroscopic observables'' introduced in Section~\ref{sec:intuition}. They also require conditions on the size of the set $A$ under consideration, in particular that its measure is not too small.

\textbf{Corollary 1} \textit{Let $A\subset I_n$ be a non-empty set. Let $f:I_n\rightarrow \mathbb{R}$ be a Lipschitz function with median $m^A_f\neq 0$ inside $A$. Moreover, we ask, \textbf{condition 1}: the set $A$ is such that $\mu_n\{A\}$ is at least $\mathcal{O}(n^{-\alpha})$, with $\alpha<1$, and \textbf{condition 2}: $\mathcal{O}(\kappa\sqrt{\ln n\,n})<\mathcal{O}(\vert m_f^A\vert)$, then}
\begin{equation}
\label{eq:corollary2}
\mu_n\left\{x\in A\,:\,\frac{\vert f(x)-m_f^A \vert}{\vert m_f^A\vert}\leq\,a_n\right\}\geq\mu_n\{A\}-b_n,
\end{equation}
\textit{with}
\begin{equation}
\label{eq:corollary_1}
\lim_{n\rightarrow\infty}a_n,\,b_n=0.
\end{equation}
\textit{If $m^A_f=0$, \textbf{condition 2} is replaced by $\kappa\sqrt{\ln n\,n}\xrightarrow[n\to\infty]0$, and the statement is,}
\begin{equation}
\label{eq:corollary2}
\mu_n\left\{x\in A\,:\,\vert f(x)-m_f^A \vert\leq\,a_n\right\}\geq\mu_n\{A\}-b_n,
\end{equation}
\textit{with vanishing $a_n$ and $b_n$ for large $n$}.

Note that \textbf{condition 1} is over the set and \textbf{condition 2} is over the observable.

We will illustrate this result with an example. Let us define the density as $\rho(x)=\frac{1}{n}\sum_{i=1}^n\,\xi_i$. In the appendix we show that this is a Lipschitz function with $\kappa^* \leq\frac{1}{n}$. Since the median on any set $A$ is $\mathcal{O}(m^A_{\rho})=1$,
\begin{equation}
\frac{\kappa\sqrt{\ln n\,n}}{{\vert m_{\rho}^A\vert}}\xrightarrow{n\to\infty}\,0.
\end{equation}
Then, if the set is not too small (at least $\mathcal{O}(n^{-\alpha})$), we can conclude that the values of the density will concentrate sharply around the median for very large $n$.

Now we will show an example in which the conditions of the corollary fail and we can thus expect concentration not to occur. Consider the observable $s_i(x)=\xi_i$, it is easy to show that in this case $\kappa^*=1$. Moreover, the function and consequently its median is of order $1$, $\mathcal{O}(\vert m_{s_i}^A\vert)=1$. Then, it is easy to check that  $a_n$ diverges for increasing $n$ in this case. We conclude that, according to our intuition, $s_i(x)$ is not a macroscopic observable.

\textbf{Corollary 2: Dynamical typicality} \textit{Let $A_R\subset I_n$ be a non-empty set conformed by states that satisfy certain constraints $R$. Let $T:I_{n}\rightarrow\, I_{n}$ define a discrete evolution rule, such that $x_{m+1}=T(x_m)$, where $m$ is the time-step index. Let us consider the set of states resulting after $m$ time-steps of the dynamics,}
\begin{equation}
A(m)=T^{m}(A(0)),
\end{equation}
\textit{with $A(0)=A_R$. If $\mu_n\{A(m)\}$ is at least $\mathcal{O}(n^{-\alpha})$, with $\alpha<1$ for all $m\geq 0$, and $\mathcal{O}(\kappa\sqrt{\ln n\,n})<\mathcal{O}(\vert m_f^{A(m)}\vert)$ for all $m\geq 0$, i.e., if $A(m)$ is large enough to fall under the hypothesis of \textbf{Corollary 1}, then,}
\begin{equation}
\mu_n\left\{x\in A(m)\,:\,\frac{\vert f(x)-m_f^{A(m)} \vert}{\vert m_f^{A(m)}\vert}\leq\,a_n(m)\right\}\geq\mu_n\{A(m)\}-b_n(m),\,\forall m\geq 0,
\end{equation}
\textit{with}
\begin{equation}
\label{eq:corollary_1}
\lim_{n\rightarrow\infty}a_n(m),\,b_n(m)=0,\,\forall m\geq 0.
\end{equation}
\textit{In words, if the dynamics does not compress the set of states below $\mathcal{O}(n^{-\alpha})$, with $\alpha<1$, then the values of all macroscopic observables will concentrate around the median for large $n$, for all $m$. It is easy to convince oneself that this implies the intuitive \textbf{Dynamical typicality} statement provided in Section~\ref{sec:intuition}}.




\subsection{Proof and comments}
First we will prove the main theorem, following closely the logic in Ref.~\cite{barvinok05_com}.
We will first define two sets,
\begin{eqnarray}
A_{+}&=&\left\{ x\in A\,:\,f(x)\geq m_f^A \right\},\\
A_{-}&=&\left\{ x\in A\,:\,f(x)\leq m_f^A \right\}.
\end{eqnarray}
From the definition of median, we have $\mu_n\{A_+\}\,,\,\mu_n\{A_-\}\geq\,\mu_n\{A\}/2$. The next step is to define the $\epsilon\sqrt{n}$-neighborhoods of these sets,
\begin{eqnarray}
A_{+}(\epsilon)&=&\left\{ x\in A\,:\,\mathrm{dist}(x,A_+)\leq \epsilon\sqrt{n} \right\},\\
A_{-}(\epsilon)&=&\left\{ x\in A\,:\,\mathrm{dist}(x,A_-)\leq \epsilon\sqrt{n} \right\},
\end{eqnarray}
where the distance of a point $x$ to a set $A$ is defined as $\mathrm{dist}(x,A)=\min_{y\in A}\mathrm{dist}(x,y)$.

Our main task is to find bounds for the measure of $A_{+}(\epsilon)$ and $A_{-}(\epsilon)$. In order to do so, we will use a fundamental concentration bound in the Boolean cube, a bound on the size of $\epsilon\sqrt{n}$-neighborhoods of a given subset $A$ in $I_n$,
\begin{equation}
\label{eq:concentration}
\mu_n\left\{ x\in I_n\,:\,\mathrm{dist}(x,A)\geq \epsilon\sqrt{n} \right\}\leq\frac{e^{-\epsilon^2}}{\mu_n\{A\}}.
\end{equation}
A proof of this result can be found in Ref.~\cite{barvinok05_com}. The result indicates that a neighborhood of moderate size $\epsilon\sqrt{n}$ of a sufficiently large subset $A$ covers almost all of $I_n$. This result can be used for our purposes by noticing that given two subsets $A,B\subset I_n$, Eq.~(\ref{eq:concentration}) implies that,
\begin{equation}
\label{eq:our_bound}
\mu_n\left\{ x\in A\,:\,\mathrm{dist}(x,B)\geq \epsilon\sqrt{n} \right\}\leq\frac{e^{-\epsilon^2}}{\mu_n\{B\}},
\end{equation}
since, clearly,
\begin{equation}
\mu_n\left\{ x\in A\,:\,\mathrm{dist}(x,B)\geq \epsilon\sqrt{n} \right\}\leq \mu_n\left\{ x\in I_n\,:\,\mathrm{dist}(x,B)\geq \epsilon\sqrt{n} \right\}.
\end{equation}

We will use Eq.~(\ref{eq:our_bound}) to bound $\mu_{n}\{A_{+}(\epsilon)\}$. First note that since
\begin{equation}
\mu_n\left\{ x\in A\,:\,\mathrm{dist}(x,A_+)\geq \epsilon\sqrt{n} \right\}+\mu_n\left\{ x\in A\,:\,\mathrm{dist}(x,A_+)< \epsilon\sqrt{n} \right\}=\mu_n\{A\},
\end{equation}
we obtain
\begin{equation}
\mu_n\left\{ x\in A\,:\,\mathrm{dist}(x,A_+)< \epsilon\sqrt{n} \right\}\geq\,\mu_n\{A\}-\frac{e^{-\epsilon^2}}{\mu_n\{A_+\}}.
\end{equation}
Now, given that $\mu_n\{A_+\}\geq\,\mu_n\{A\}/2$, we conclude that,
\begin{equation}
\mu_n\{A_+(\epsilon)\}\geq\,\mu_n\{A\}-\frac{2e^{-\epsilon^2}}{\mu_n\{A\}}.
\end{equation}
In a completely analogous way we can show that,
\begin{equation}
\mu_n\{A_-(\epsilon)\}\geq\,\mu_n\{A\}-\frac{2e^{-\epsilon^2}}{\mu_n\{A\}}.
\end{equation}

To proceed with our proof, we are interested in the intersection $A_+(\epsilon)\cap A_-(\epsilon)$. Clearly, this intersection exists, since $A_+$ and $A_{-}$ both cover at least a half of $A$, and $A_{+}(\epsilon),\,A_{-}(\epsilon)$ are even bigger than them. We can calculate the measure of the intersection as,
\begin{equation}
\mu_n\left\{A_+(\epsilon)\cap A_-(\epsilon)\right\}=\left(\mu_n\left\{A_+(\epsilon)\right\}-\frac{\mu_n\left\{A\right\}}{2}\right)+\left(\mu_n\left\{A_-(\epsilon)\right\}-\frac{\mu_n\left\{A\right\}}{2}\right),
\end{equation}
which implies that,
\begin{equation}
\mu_n\left\{A_+(\epsilon)\cap A_-(\epsilon)\right\}\geq \mu_n\left\{A\right\}-\frac{4e^{-\epsilon^2}}{\mu_n\left\{A\right\}}.
\end{equation}
Now, for every $x\in A_+(\epsilon)\cap A_-(\epsilon)$, since $f(x)$ is a Lipschitz function, we have that $\vert f(x)-m_f^A \vert\leq \kappa\epsilon\sqrt{n}$, which completes the proof.

To prove corollary 1 we have to choose $\epsilon=\sqrt{\ln n}$ in Eq.~(\ref{eq:main_result}) to obtain Eq.~(\ref{eq:corollary2}) with $a_n=\frac{\kappa\sqrt{\ln n\,n}}{{\vert m_f^A\vert}}$ and $b_n=\frac{1}{n\,\mu_n\{A\}}$. In the case $m_f^A=0$, $a_n=\kappa\sqrt{\ln n\,n}$. Corollary 2 is automatically proved once corollary 1 is established.

\section{Application to one-dimensional cellular automata}
\label{sec:application}

In this section we will develop an application of the results presented in the previous sections in a concrete example. We will work with a 1D chain of $n$ binary sites. The set $A$ will be the set containing all configurations with density $\rho=\frac{1}{n}\sum_{i}\xi_i=1/2$:
\begin{equation}
A=\left\{x\in I_n : \rho(x)=\frac{1}{2}\right\}.
\end{equation}
The size of $A$ is $\binom{n}{n/2}\simeq \sqrt{\frac{2}{\pi}}\,n^{-1/2}\,2^n$, which implies, $\mu_n\{A\}\simeq \sqrt{\frac{2}{\pi}}\,n^{-1/2}$. This set is then large enough for the Corollary 1 to be of relevance, i.e., \textbf{condition 1} is fulfilled.

Let us introduce more observables into the description of the system. We will first define the density of domain walls $\rho_{\vert}$ as the number of occurrences of the sequences $10$ or $01$ in the configuration. This  observable can be written as,
\begin{equation}
\rho_{\vert}=\frac{1}{n}\sum_{i=1}^{n}\,\vert \xi_i-\xi_{i+1} \vert.
\end{equation}
In this one dimensional system, the density of domain walls can be related to the Ising energy $E=-J \sum_{i=1}^{n}S_i S_{i+1}$, where $S_i=2\xi_i-1$ is the "spin" variable associated with site $i$, through $E=-J n (1- 2 \rho_{\vert})$.
We will also introduce a correlation function, $C_2(r)$~\cite{wolfram83_ca_stat-mech}, defined as,
\begin{equation}
C_2(r)=\frac{1}{n}\sum_{i=1}^{n}\,S_i S_{i+r}-\left(\frac{1}{n}\sum_{i=1}^{n}\,S_i \right)\left(\frac{1}{n}\sum_{i=1}^{n}\,S_{i+r} \right).
\end{equation}

\begin{figure}[!h]
\begin{center}
\includegraphics[scale=0.7]{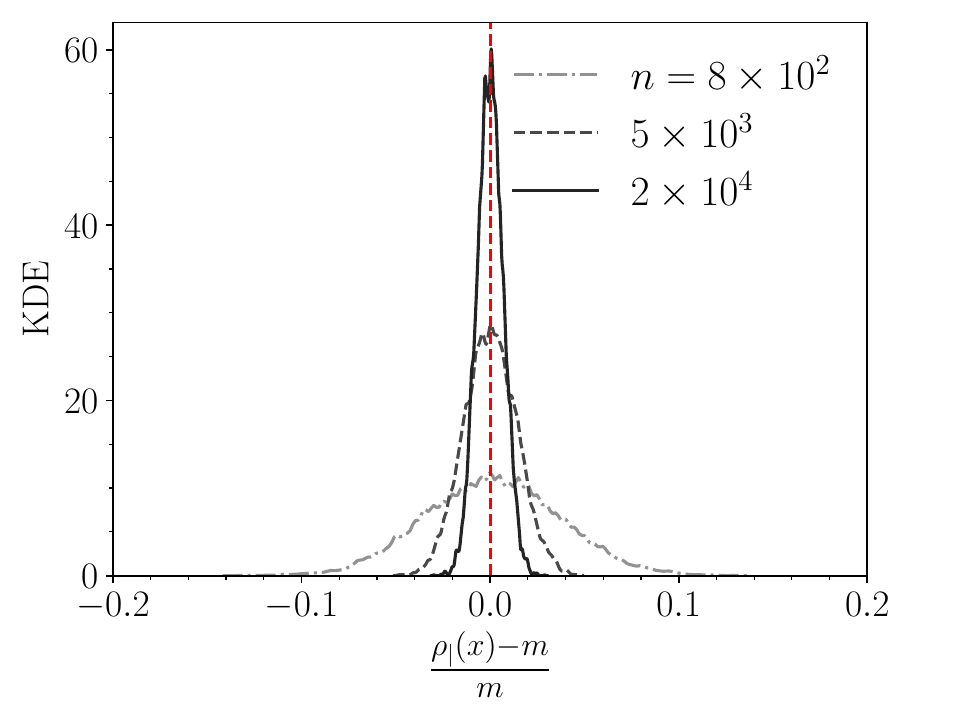}
\end{center}
\caption{\small {\bf Static typicality}
Kernel density estimation of the probability density for the relative deviation of the domain wall density $\rho_{\vert}$ from the median of a sample of $10^4$ random configurations with density $\rho=1/2$ for different system sizes. As the system size increases, the probability density becomes sharply peaked around zero, which is consistent with our analytical result.}
\label{fig:fig1}
\end{figure}

The domain wall density $\rho_{\vert}$ and the correlation function $C_2(r)$ are Lipschitz functions (see Appendix). Moreover, it is easy to check that these observables fulfill \textbf{condition 2}, $\mathcal{O}(\kappa\sqrt{\ln n\,n})<\mathcal{O}(\vert m_f^A\vert)$, i.e., they are macroscopic observables according to our definition. Then, corollary 2 applies in this set up, which implies that, for large $n$, $\rho_{\vert}$ and $C_2(r)$ concentrate sharply around their medians inside $A$. In other words, for the vast majority of states in $A$, the value of these observables should be approximately the same, which implies that their distributions inside $A$ will show enormously sharp peaks. And this will be true also for any other observable fulfilling \textbf{condition 2}. In order to illustrate this, in Fig.~\ref{fig:fig1} we show the kernel density estimation (KDE) of the probability density for the relative deviation of the domain wall density $\rho_{\vert}$ from the median inside a sample of $10^4$ random configurations with density $\rho=1/2$ for different system sizes. It is clearly visible that as the size of the system increases, the distribution becomes more sharply peaked around the median, which is consistent with our analytical result.

\begin{figure}[!h]
\begin{center}
\includegraphics[scale=0.5]{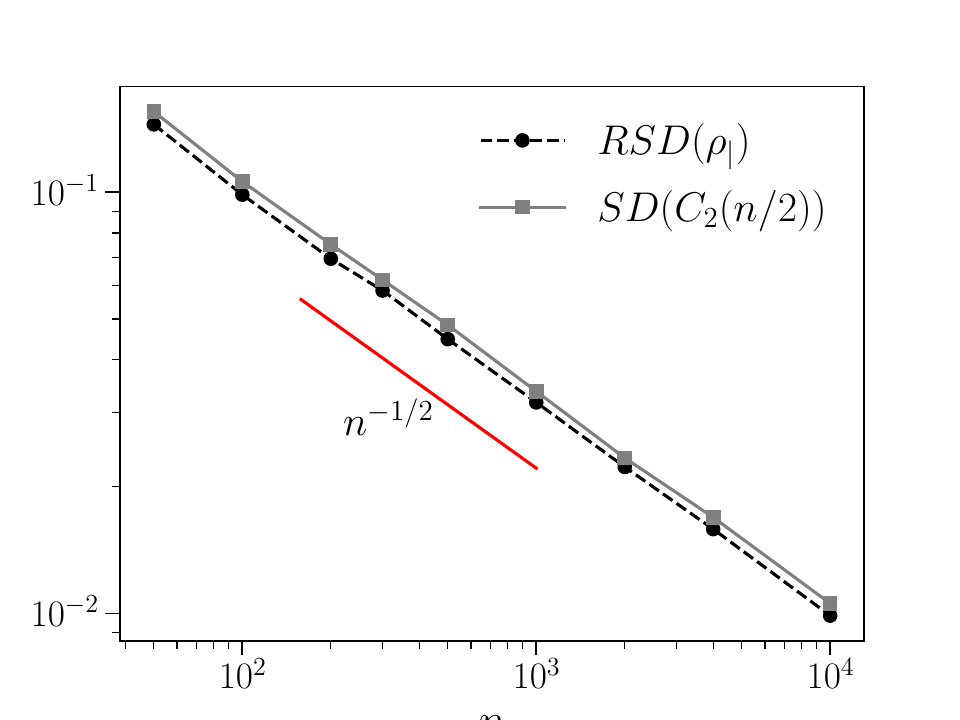}
\end{center}
\caption{\small {\bf Static typicality}
RSD of $\rho_{vert}$ and $SD$ of $C_2(n/2)$ with respect to the median as a function of system size for a sample of $10^4$ random states with $\rho=1/2$. Note the double logarithmic scale.}
\label{fig:fig2}
\end{figure}

Another way to visualize the typicality phenomenon is to study the variance around the median for a sample of states inside the set $A$ as $n$ increases. Such variance can be defined as,
\begin{equation}
\label{eq:dispersion}
\sigma^2_{m}=\frac{1}{\vert S \vert}\sum_{x\in S}\,\vert f(x)-m\vert^2,
\end{equation}
where $S$ is the sample and $m$ is the median of $f$ inside $S$. If $S$ is large enough and samples the set $A$ unbiasedly, the variance of the sample will be a good estimator of the variance in $A$. In Fig.~\ref{fig:fig2} we show the relative standard deviation $RSD=\sqrt{\sigma^2_{m}/m^2}$ of $\rho_{\vert}$ and the standard deviation $SD=\sigma_{m}$ for $C_2$ evaluated at $r=n/2$ as a function of system size. We choose the standard deviation, and not the relative standard deviation, for $C_2(n/2)$ because the median is very close to zero in this case. As can be clearly seen from the plot, both dispersion measures decay as a power law with exponent $-1/2$ as $n$ increases. Although this result is completely consistent with our analytical bound, we were not able to recover analytically the power law decay from our result.

In order to address dynamical typicality in this one dimensional set up, we need to define a discrete dynamical rule. We will focus on the elementary cellular automata (ECA) rules. In an ECA, the rule which determines the state of the site in the next time step depends on the current state of the site and its two immediate neighbors. There are only $2^{2^3}=256$ different dynamical rules allowed under this scheme, which can be labeled using Wolfram's numbering system, which assigns an integer between $0$ and $255$ to each rule~\cite{wolfram83_ca_stat-mech}. There are some basic transformations that can be applied to ECA rules, namely, reflection (left-right inversion) and conjugation ($1\leftrightarrow 0$ exchange). Through these transformations it is possible to establish $88$ equivalence classes which cover the complete rule space. Moreover, ECA rules can be classified into four classes according to the behavior of the system starting from a random state~\cite{wolfram02_nks_book}. Class $1$ and $2$ show very simple dynamics either quickly reaching a uniform state (class $1$) or showing a repetitive pattern (class $2$). On the other hand, class $3$ shows apparently random or noisy dynamical patterns, while class $4$ rules produce a more complex and structured dynamics. A finite cellular automata has $2^n$ possible configurations, which implies that the dynamics should become periodic after at most $2^n$ time steps. After a transient, the ECA dynamics must enter a cycle. Simple cellular automata (class 1 and class 2) yield short cycles with only a few configurations, while complex cellular automata (class $3$ and $4$) may yield much longer cycles. In particular, complex cellular automata may yield cycles whose length increases without bound with $n$.

\begin{figure}[!h]
\begin{center}
\includegraphics[scale=0.5]{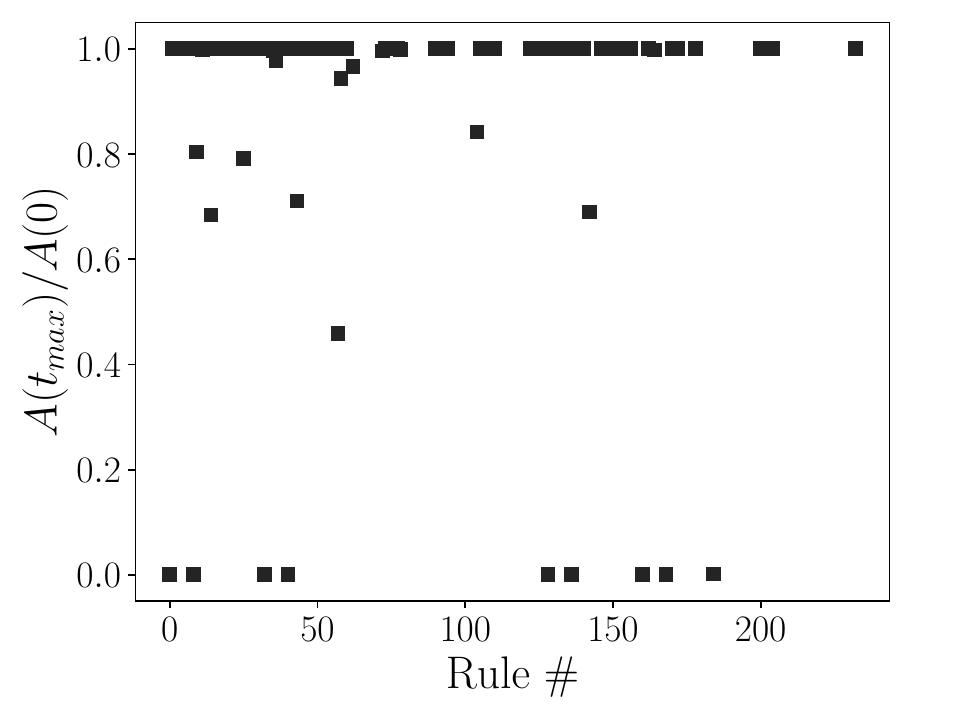}
\end{center}
\caption{\small {\bf Local irreversibility of ECA rules}
Set size ratio between the final set $A(t_max)$ and the initial set of states $A(0)$ for each one of the $88$ representative ECA rules. In all cases $\vert A(0) \vert=10^3$. All states are random states with density $\rho=1/2$.}
\label{fig:fig3}
\end{figure}

A relevant feature of ECAs in our context is its ``local irreversibility'', which means that, during the dynamics, ECA rules can coalesce two different states into the same state. In other words, every state has unique descendants, but not unique predecessors. In particular, there are special states that can never appear as a result of dynamic evolution, and can only be considered as initial conditions for the dynamics. The irreversible or dissipative nature of ECA clearly induces a reduction of the size of the set during the dynamics. In Fig.~\ref{fig:fig3}, we study local irreversibility for the $88$ representative ECA rules. Taking a set of $10^3$ random states with density $\rho=1/2$ as initial conditions, for each rule we plot the quotient between the initial number of different states and the number of different states after $100$ steps of evolution for a system with $n=100$. We see that for the majority of rules the fraction is very close or exactly $1$. This ensures that the majority of ECA rules are within the conditions of the dynamical typicality theorem, corollary 2. However there are two kinds of rules for which an appreciable number of states are merged during the dynamics. The trivial case are the rules in class 1, which, after a short transient, evolves any state into a uniform state, i.e., all states collapse into a single state. On the other hand, there are some class 2 rules which merge a considerable number of states during the dynamics, producing the points in the intermediate region between $0$ and $1$ in Fig.~\ref{fig:fig3}. However, not all class 2 rules show this behavior. For example, Rule 4, a class 2 representative, preserves the number of different states after $100$ steps of evolution. Finally, a special case is that of Rule $184$ which, according to the results, evolves all initial states into one out of two possible final states. This is due to the fact that, for the case of $\rho=1/2$, Rule $184$ collapses all states into an alternating sequence of $1$s and $0$s~\cite{li92_non-local_CA}, leaving two possible states according to which is the first digit of the chain.

\begin{figure}[!h]
\begin{center}
\includegraphics[scale=0.5]{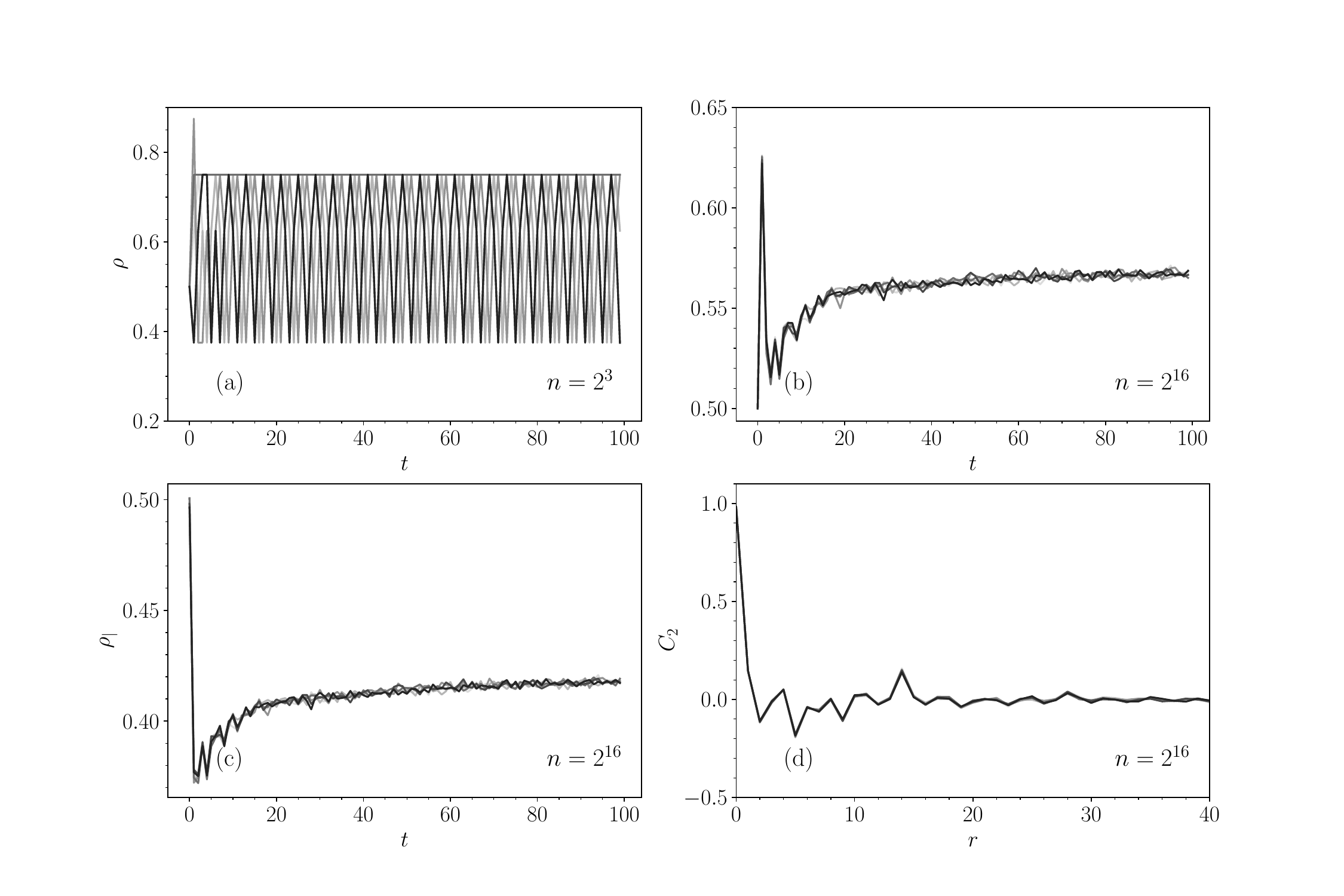}
\end{center}
\caption{\small {\bf Dynamic typicality}
Dynamics of macroscopic observables for random initial conditions with $\rho=1/2$, evolved according to ECA rule $110$ for system size $n=2^3$ and $n=2^16$.}
\label{fig:fig4}
\end{figure}

In order to illustrate dynamical typicality, in Fig.~\ref{fig:fig4} we analyze the time evolution of several observables in a small system ($n=2^3$) and a larger system ($n=2^{16}$) for different initial conditions for ECA Rule $110$ (class $4$). The initial conditions are random states with density $\rho=1/2$. It is clearly visible that for the larger system, the trajectories of all the considered observables tend to collapse on the same curve, while for the smaller system this is not true. This is exactly the type of behavior that we expected based on our analytical result on dynamical typicality, since the dynamics does not shrink the size of the initial set (corollary 2), the set from which the initial conditions were drawn is large enough (condition 1 of corollary 1) and the observables are macroscopic (condition 2 of corollary 1). It is important to note that, for the smaller system, cyclic recurrence can dominate the dynamics, inducing deviations from typicality.

We would like to mention that there could be special initial conditions which do not exhibit typical dynamical behavior. That would be the case for some special periodic initial conditions that evolve into a uniform state for ECA rules $110$ and $126$. However, our analytical results ensure that such atypical initial conditions have zero measure in the limit of large $n$.

\begin{figure}[!h]
\begin{center}
\includegraphics[scale=0.5]{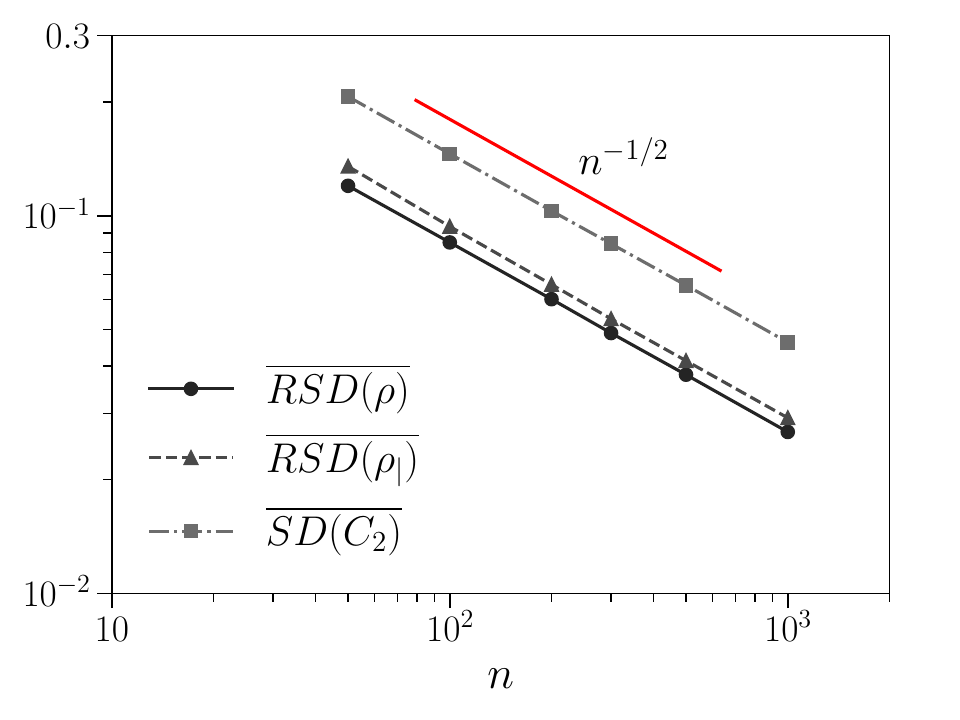}
\end{center}
\caption{\small {\bf Dynamic typicality}
Time average of RSD for $\rho$ and $\rho_{vert}$ and $SD$ for $C_2(n/2)$ with respect to the median as a function of system size for a sample of $5\times10^3$ random initial conditions with $\rho=1/2$ evolved during $t_{max}=50$ time steps according to ECA rule $110$. Note the double logarithmic scale.}
\label{fig:fig5}
\end{figure}

In Fig.~\ref{fig:fig5} we show the time average of the relative standard deviation for $\rho$ and $\rho_{\vert}$ and the time average of the standard deviation for $C_2$ for a set of $5\times 10^3$ random initial conditions with density $1/2$ evolving with rule $110$. Again the decrease of the standard deviation with increasing system size is consistent with the dynamical typicality theorem. Moreover, the fact that the decay follows the same law as in the static case supports the picture revealed by our analytical results according to which static and dynamic typicality are basically the same phenomenon.

\begin{figure}[!h]
\begin{center}
\includegraphics[scale=0.5]{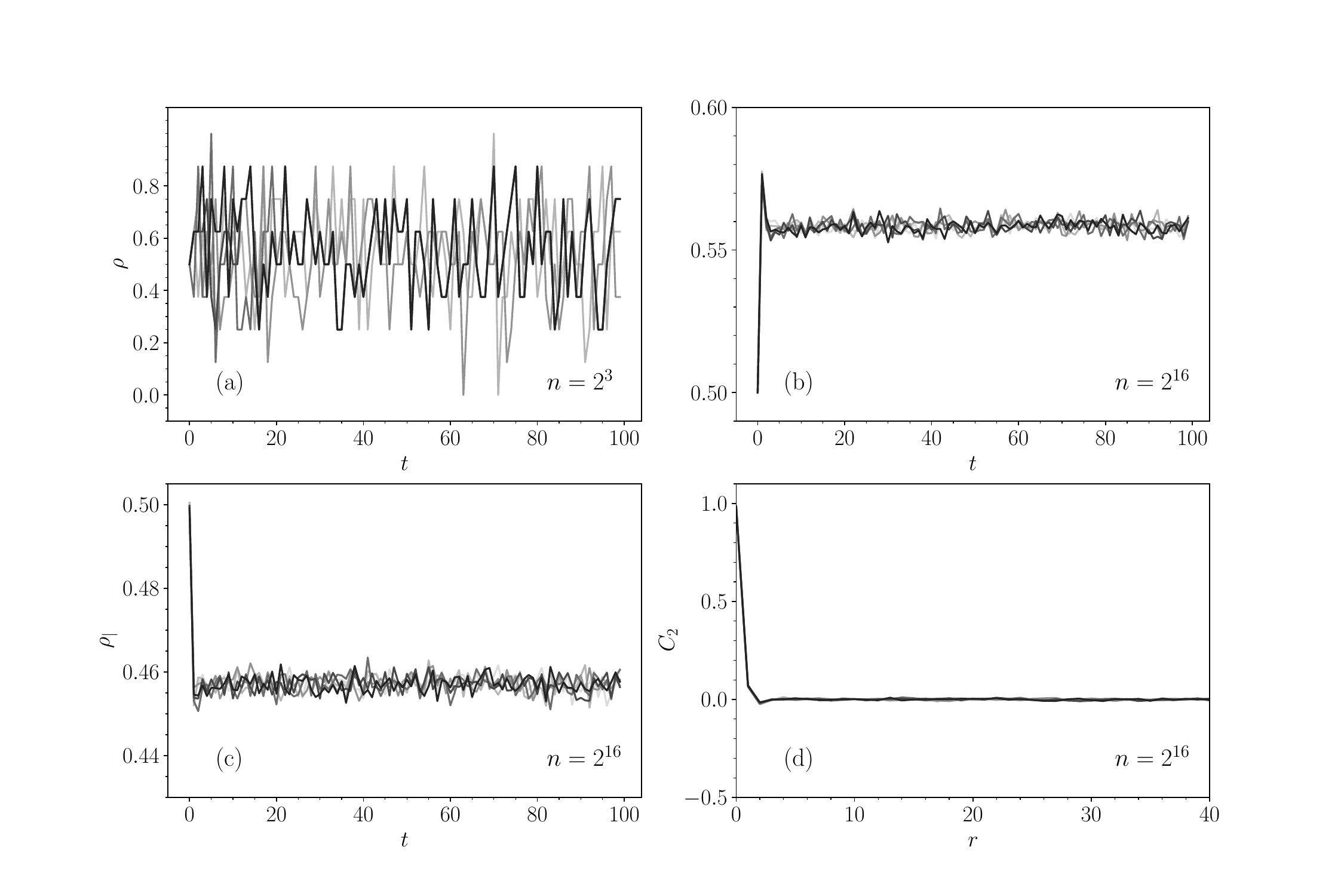}
\end{center}
\caption{\small {\bf Dynamic typicality in PCA}
Dynamics of macroscopic observables for random initial conditions with $\rho=1/2$, evolved according to the PCA rule described in the main text for $p=0.8$ for system size $n=2^3$ and $n=2^16$.}
\label{fig:fig6}
\end{figure}

Finally, we will analyze dynamical typicality in the case of a dynamic rule with a random component, a probabilistic cellular automaton (PCA). In particular, we will analyze a probabilistic variation of the deterministic ECA rule $110$. The probabilistic rule is obtained from the deterministic one by prescribing a successor state $\xi(t+1)=1$ with probability $p$ ($\xi(t+1)=0$ with probability $1-p$) whenever the deterministic rule prescribes a successor state $\xi(t+1)=1$, and conversely for the case in which the deterministic rule prescribes a successor state $\xi(t+1)=0$. We find that the fraction of final set size to initial set size is exactly $1$, for an initial set with $10^3$ random initial conditions with density $\rho=1/2$, and $n=100$ after $100$ time steps. This implies that this probabilistic rule fulfills the conditions of corollary 2 and we can expect dynamical typicality. In fact, in Fig.~\ref{fig:fig6} it is clearly visible the collapse of the dynamical trajectories for the considered observables for large system sizes. This result, consistent with our analytical findings, shows that dynamical typicality does not depend on the nature of the dynamics (deterministic or random) as long as the dynamics is not strongly irreversible or dissipative.

We would like to note that the analytical results presented in this paper provide \textit{sufficient} conditions for the presence of typicality, which implies that there might be situations outside the conditions of our theorems which exhibit this behavior. In particular, we have checked that there is dynamical typicality for sets which are smaller than those fulfilling condition 1 of corollary 1. For example, the set $B=\{ x\in I_n : \rho(x)=1/4 \}$, whose measure is \textit{exponentially} small in $n$, displays static typicality, and also dynamic typicality, if we use the states in such set as initial conditions for the ECA rules. However, we have also checked that when the measure of the set is very small, such as sets of states with very low or very high density, static and dynamic typicality is not present.

\section{Conclusions and outlook}
\label{sec:conclusions}

We have obtained analytical results regarding the behavior of macroscopic observables in large subsets of the Boolean cube. We have shown that such macroscopic observables are sharply concentrated around the median inside the given set for sufficiently large $n$. Moreover, such result allow us to infer that \textit{any} dynamical rule which is not strongly dissipative, such that it does not shrink abruptly the size of the set from where the initial conditions were drawn, will display dynamical typicality. This constitutes, as far as we know, the first example of dynamical typicality in a classical system, in contrast with the quantum mechanical case in which typicality has been studied extensively. It is important to note that in the quantum case even microscopic observables, like the quantum expectation value of a single spin in a spin chain, show typical dynamics for large system sizes~\cite{bartsch09_dyn_typ_first}. This is at variance with what we found, since in classical binary state systems, microscopic observables can show large fluctuations even for large systems.

It would be interesting to try to improve the analytical bounds presented in this paper, in order to prove typicality in sets whose measure is exponentially small in $n$. Moreover, it would be important to extend this kind of results to systems with variables which can take more than two states and, ideally, reach the limit of continuous variables, such as those present in classical Hamiltonian systems.

\newpage

\bibliographystyle{phaip}

\appendix

\appendixpage



In this appendix we will bound $\kappa^*$ for several observables appearing in the main text. We will start with the density, $\rho(x)$,
\begin{equation}
\vert \rho(x)-\rho(y)\vert\leq \frac{1}{n}\sum_i\vert \xi_i-\eta_i\vert=\frac{1}{n}\mathrm{dist}(x,y),
\end{equation}
which implies $\kappa^*\leq\frac{1}{n}$.

Next we will address the domain wall density, $\rho_{\vert}(x)$,
\begin{equation}
\label{eq:kappa_dw}
\vert \rho_{\vert}(x)-\rho_{\vert}(y)\vert\leq \frac{1}{n}\sum_i\vert \vert \xi_i-\xi_{i+1}\vert-\vert \eta_i-\eta_{i+1}\vert \vert.
\end{equation}
Now we use the following inequality, valid for any 4-tuple of binary variables $\left( \alpha,\beta,\gamma,\delta \right)$,
\begin{equation}
\vert \vert \alpha-\beta \vert - \vert \gamma-\delta \vert\vert\leq \vert \vert \alpha-\gamma \vert + \vert \beta-\delta \vert\vert.
\end{equation}
Applying this to Eq.~(\ref{eq:kappa_dw}) we obtain,
\begin{equation}
\vert \rho_{\vert}(x)-\rho_{\vert}(y)\vert\leq \frac{1}{n}\left\{ \sum_i \vert \xi_i-\eta_{i}\vert+\sum_i \vert \xi_{i+1}-\eta_{i+1}\vert \right\}=\frac{2}{n}\mathrm{dist}(x,y),
\end{equation}
which implies $\kappa^*\leq\frac{2}{n}$.

Now we turn to the correlation function $C_2(x;r)$,
\begin{eqnarray}
\label{eq:c2_kappa}
\vert C_2(x;r)-C_2(y;r)\vert &\leq& \bigg\vert \frac{1}{n} \bigg[ \sum_i (2\xi_i -1)(2\xi_{i+r} -1) - (2\eta_i -1)(2\eta_{i+r} -1) \bigg]\\
\nonumber &+& \bigg[ \bigg( \frac{1}{n}\sum (2\eta_i-1) \bigg)\bigg( \frac{1}{n}\sum (2\eta_{i+r}-1) \bigg) -\bigg( \frac{1}{n}\sum (2\xi_i-1) \bigg)\bigg( \frac{1}{n}\sum (2\xi_{i+r}-1) \bigg) \bigg] \bigg\vert.
\end{eqnarray}
First, we bound the first term inside the absolute value bars in the r.h.s. of Eq.~(\ref{eq:c2_kappa}), which can be written as,
\begin{eqnarray}
\nonumber \frac{1}{n}\bigg\vert \bigg[ \sum_i (2\xi_i -1)(2\xi_{i+r} -1) - (2\eta_i -1)(2\eta_{i+r} -1) \bigg]\bigg\vert&\leq&\frac{1}{n}\sum_i\bigg(  4\vert \xi_i \xi_{i+r} - \eta_i \eta_{i+r} \vert+2 \vert \eta_i-\xi_i\vert+2\vert \eta_{i+r}-\xi_{i+r}\vert \bigg)\\
\nonumber &\leq& \frac{1}{n}\sum_i\bigg(  4\vert \xi_i - \eta_i\vert+4\vert \xi_{i+r}- \eta_{i+r} \vert+2 \vert \eta_i-\xi_i\vert+2\vert \eta_{i+r}-\xi_{i+r}\vert \bigg)\\
&=&\frac{12}{n} \mathrm{dist}(x,y).
\end{eqnarray}
Where, in order to obtain the last inequality, we applied the following inequality, valid for any 4-tuple of binary variables $\left( \alpha,\beta,\gamma,\delta \right)$,
\begin{equation}
\vert \alpha\beta- \gamma\delta \vert\leq \vert\alpha-\gamma \vert + \vert \beta-\delta \vert.
\end{equation}
The second term inside the absolute value bars in Eq.~(\ref{eq:c2_kappa}) can also be bound using the same inequality, obtaining,
\begin{eqnarray}
\nonumber \bigg\vert\bigg[ \bigg( \frac{1}{n}\sum (2\eta_i-1) \bigg)\bigg( \frac{1}{n}\sum (2\eta_{i+r}-1) \bigg) -\bigg( \frac{1}{n}\sum (2\xi_i-1) \bigg)\bigg( \frac{1}{n}\sum (2\xi_{i+r}-1) \bigg) \bigg] \bigg\vert &\leq& \frac{6}{n} \sum_i \vert \xi_i-\eta_i\vert +\frac{6}{n} \sum_i \vert \xi_{i+r}-\eta_{i+r}\vert\\
&=& \frac{12}{n}\mathrm{dist}(x,y).
\end{eqnarray}
Adding the contributions coming from both terms, we find,
\begin{equation}
\vert C_2(x;r)-C_2(y;r)\vert \leq \frac{24}{n}\mathrm{dist}(x,y),
\end{equation}
which implies, $\kappa^*\leq \frac{24}{n}$.

Finally, we will show that the Ising energy $E(x)$ is also a Lipschitz function,
\begin{eqnarray}
\nonumber \vert E(x)-E(y)\vert&=&J\bigg\vert \sum_i \left[ 4(\xi_i \xi_{i+1}-\eta_{i}\eta_{i+1})-2(\xi_i-\eta_i)-2(\xi_{i+1}-\eta_{i+1})\right] \bigg\vert\\
&\leq& 12 J \mathrm{dist}(x,y),
\end{eqnarray}
which implies, $\kappa^*\leq 12 J$. In this case $\kappa^*$ is order $1$ but, since $m^A_E$ is order $n$, condition 2 of corollary 1, $\mathcal{O}(\kappa\sqrt{n\ln n})<\mathcal{O}(m^A_E)$ still holds and $E(x)$ can be considered a macroscopic observable under our definition.

\end{document}